\documentclass[10pt,twocolumn]{article}
\usepackage{latexsym}
\usepackage{amssymb,amsbsy,amsmath,amsfonts,amssymb,amscd}
\usepackage{graphicx,color}
\usepackage{float,url}
\usepackage{cancel}
\usepackage[colorlinks=true]{hyperref}
\usepackage{cleveref}
\usepackage[font=it,font=small,labelfont=sc]{caption}
\usepackage{subcaption}
\usepackage[]{inputenc}
\usepackage{dblfloatfix}
\usepackage{gensymb}
\usepackage{multicol}
\usepackage{epstopdf}
\usepackage{comment}

\newcommand{\commentout}[1]{}

\newcommand {\e}  {\varepsilon}

\newcommand {\Chi} {{\bf \raise 2pt \hbox{$\chi$}} }
\newcommand{\bea} {\begin{array}{rl}}
\newcommand{\eea} {\end{array}}
\newcommand{\bepa}{\left\{ \begin{array}{l}}
\newcommand{\eepa} {\end{array}\right.}
\renewcommand{\d}{\text{d}}

\newtheorem{theorem}{Theorem}[section]

\newtheorem{remark}[theorem]{Remark}

\newcommand{\qed}{{ \hfill
                       {\unskip\kern 6pt\penalty 500 \raise -2pt\hbox{\vrule\vbox to 6pt{\hrule width 6pt
                       \vfill\hrule}\vrule} \par}   }}
\title{A chemotaxis-based explanation of spheroid formation in 3D cultures of breast cancer cells}

\author{Federica Bubba$^{1*}$ \and Camille Pouchol$^{1*}$ \and Nathalie Ferrand$^{2}$ \and Guillaume Vidal$^{3}$ \and Luis Almeida$^{1,4}$ \and Beno\^it Perthame$^{1}$ \and Mich\`ele Sabbah$^{2}$}
\date{\today}

\begin{document}

\pagestyle{plain}
\pagenumbering{arabic}

\twocolumn[
\begin{@twocolumnfalse}
\maketitle
{\centering
$^{1}$ Sorbonne Universit\'{e}, CNRS, Universit\'{e} Paris-Diderot SPC, Inria, Laboratoire Jacques-Louis Lions, 4 pl. Jussieu, 75005 Paris, France.\\
\noindent $^{2}$ Sorbonne Universit\'{e}, INSERM, Laboratoire de Biologie du Cancer et Th\'{e}rapeutique, Centre de Recherche Saint-Antoine, 75012 Paris, France.\\
\noindent $^{3}$ CELENYS, Biopolis 2, 75 route de Lyons-la-for\^et, 76000 Rouen, France.\\
\noindent $^{4}$ Corresponding author: almeida@ljll.math.upmc.fr\\
\noindent $^*$ This authors contributed equally to this work.
\par%
}

\begin{abstract}
Three-dimensional cultures of cells are gaining popularity as an \textit{in vitro} improvement over 2D Petri dishes. In many such experiments, cells have been found to organize in aggregates. We present new results of three-dimensional \textit{in vitro} cultures of breast cancer cells exhibiting patterns. Understanding their formation is of particular interest in the context of cancer since metastases have been shown to be created by cells moving in clusters. In this paper, we propose that the main mechanism which leads to the emergence of patterns is chemotaxis, \textit{i.e.}, oriented movement of cells towards high concentration zones of a signal emitted by the cells themselves. Studying a Keller-Segel PDE system to model chemotactical auto-organization of cells, we prove that it is subject to Turing instability under a time-dependent condition. This result is illustrated by two-dimensional simulations of the model showing spheroidal patterns. They are qualitatively compared to the biological results and their variability is discussed both theoretically and numerically.
\end{abstract}
 \end{@twocolumnfalse}
]

\vskip .7cm

\noindent{\makebox[1in]\hrulefill}\newline

%%%%%%%%%%%%%%%%%%%%%%%%%%%%%%%%%%%%%%%%%%%%
\section*{Introduction}
\label{sec:intro}
%-------------------------------------------
%%%%%%%%%%%%%%%%%%%%%%%%%%%%%%%%%%%%%%%%%%%%

In breast cancer, the majority of deaths are not due to the primary tumor itself but are the result of the ability of cancer cells to migrate and colonize other organs in the body, \textit{i.e.}, to form metastases~\cite{Nguyen2009}. Even when they are few, highly motile tumor cells are able to move and spread throughout the entire body, but it is now well established that cancer cells creating metastases typically move in clusters and not alone~\cite{Aceto2014, Hong2016}. Moreover, organs most usually affected by breast cancer metastases, such as lungs, bones and liver, produce proteins which attract chemokine receptors in cancerous cells~\cite{Mueller2001}. The migrative properties of cells and their chemotactic abilities, \textit{i.e}, their directing movement towards zones of high concentration of certain chemical stimuli, are thus key when it comes to understanding the development of metastases in breast cancer.
 
Over the past decades, the \textit{in vitro} investigation of how cells move and organize in the ECM has developed thanks to the design of 3D structures mimicking the ECM, see~\cite{Haycock2011} for a review of the different engineering techniques. The typical behavior  of cells cultured inside these is aggregation~\cite{Lee2008}, creating patterns whose characteristics depend on the cell line~\cite{Kenny2007, Singh2016}. Cells aggregated into clusters might have a selective advantage over single cells thanks to their ability to escape the immune response, facilitating the extravasation, while it might prevent anoikis~\cite{Friedl2003, Jurasz2004, Zhao2010}.

In this paper, we are interested in determining the main biological phenomenon responsible for their formations through mathematical modeling. Models for the aggregative behavior of cells in the extracellular matrix (ECM) are well developed and belong to two main classes: discrete (or agent-based) models, where each cell is represented individually~\cite{Drasdo2005, Schluter2012, Zhou2006}, or continuous models typically based on ordinary differential equations (ODEs) or partial differential equations (PDEs)~\cite{Anderson2005, Painter2010}. The latter can be either phenomenological or derived from physical or chemical laws as outlined in~\cite{Painter2009}, where patterning is obtained but requires anisotropy assumptions on the ECM. 
Patterns are also reported in 2D Petri dishes in various cases such as buds created by glioblastoma cells. In~\cite{Agosti}, such experimental results are presented together with a PDE mechanical model which is shown to numerically reproduce aggregates. 

To the best of our knowledge, however, works on the formation of patterns by cells in the ECM or in artificial 3D structures do not incorporate chemical environmental cues inducing chemotaxis. In a deterministic setting, the latter phenomenon is commonly modeled thanks to the Keller-Segel system~\cite{KellerSegel1970} or one of its numerous generalizations \cite{Hillen2009}. Originally employed to model spatial patterns in bacteria populations, this model has been shown to be a rich tool for the modeling of self-organization phenomena~\cite{Painter2018}.

This work provides new experimental results of spheroidal aggregates created by cancerous cells in a 3D hydrogel. It proposes a chemotaxis-based explanation of patterns through theoretical and numerical analysis of Turing instabilities exhibited by a Keller-Segel-type PDE model.

The outline of the paper is as follows. In~\Cref{sec:experiments}, we present the experimental results and the main features of their PDE-based counterpart. In~\Cref{sec:model}, we introduce the full model, whose linear stability analysis is performed in \Cref{sec:analysis}. Numerical simulations of the model are given in~\Cref{sec:simulations}, and we conclude by discussing the results and possible improvements in~\Cref{sec:conclusions}.

\section{From experimental to modeling results}
\label{sec:experiments}

\begin{figure}[h]
	\centering
	\begin{subfigure}[]{0.27\textwidth}
		\includegraphics[width=\textwidth]{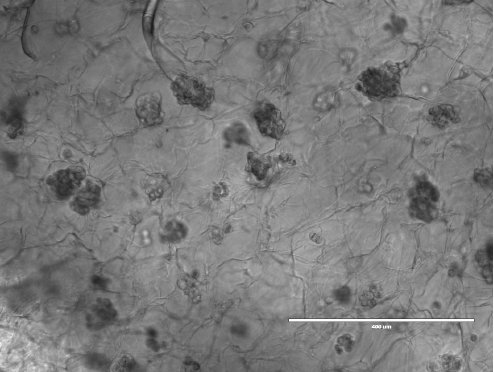}
		\caption{}
	\end{subfigure}
	\begin{subfigure}[]{0.3\textwidth}
		\includegraphics[width=\textwidth]{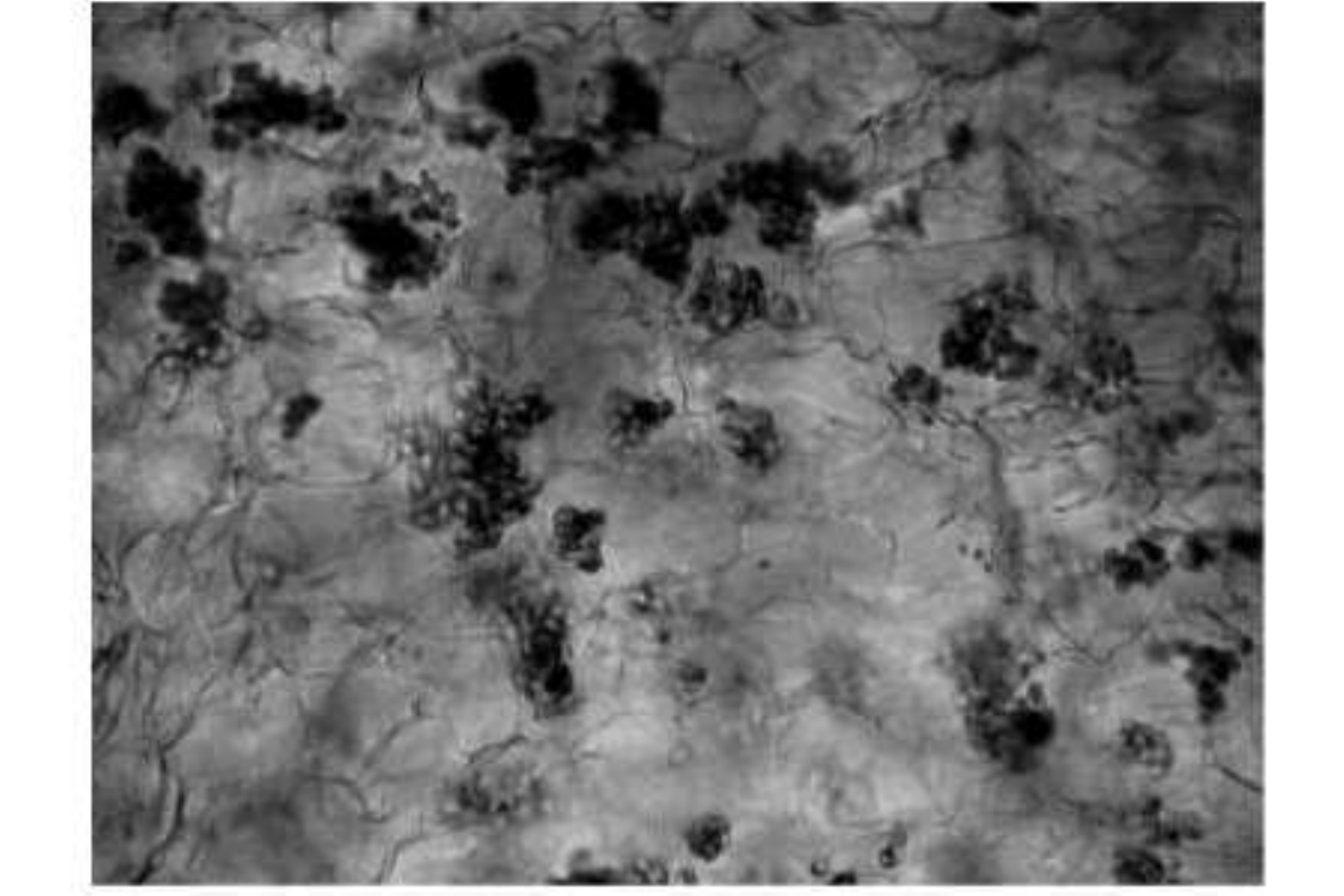}
		\caption{}
	\end{subfigure}
	\\[-2mm]
	\caption{2D image of spheroids in the hydrogel, formed by MCF7 cells (upper panel) and by cells from MCF7-sh-wisp2 (lower panel), in both cases for $75\,000$ seeded cells and after $4$ days of culture.}
	\label{Spheroids}
\end{figure}

In this Section, we report new results of experiments exhibiting spheroidal patterns, specifically for breast cancer cells growing in a 3D cylindrical hydrogel. Cells are put at the top of the structure and, after spreading uniformly, have typically formed spheroids no later than the day $4$ (D4) of culture. This is true both for an epithelial cell line (MCF7) and a more mesenchymal and invasive one (MCF7-sh-wisp2), although patterns appear to be more regular in the first cell line and exhibit more elongated shapes with geometric variability in the second, see Figure \ref{Spheroids}.

\begin{figure*}[h!!]
	\centering
	\begin{subfigure}[]{0.3\textwidth}
		\includegraphics[scale=0.45]{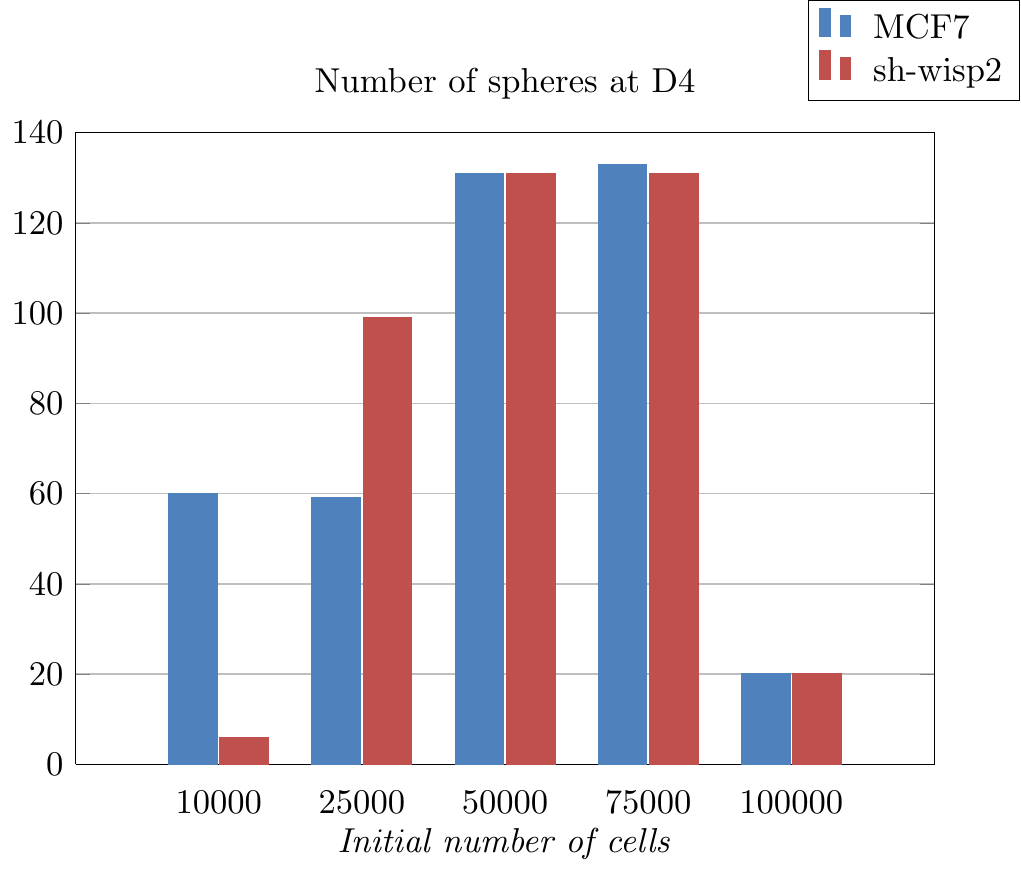}
		\caption{}
	\end{subfigure}
	\begin{subfigure}[]{0.3\textwidth}
		\includegraphics[scale=0.45]{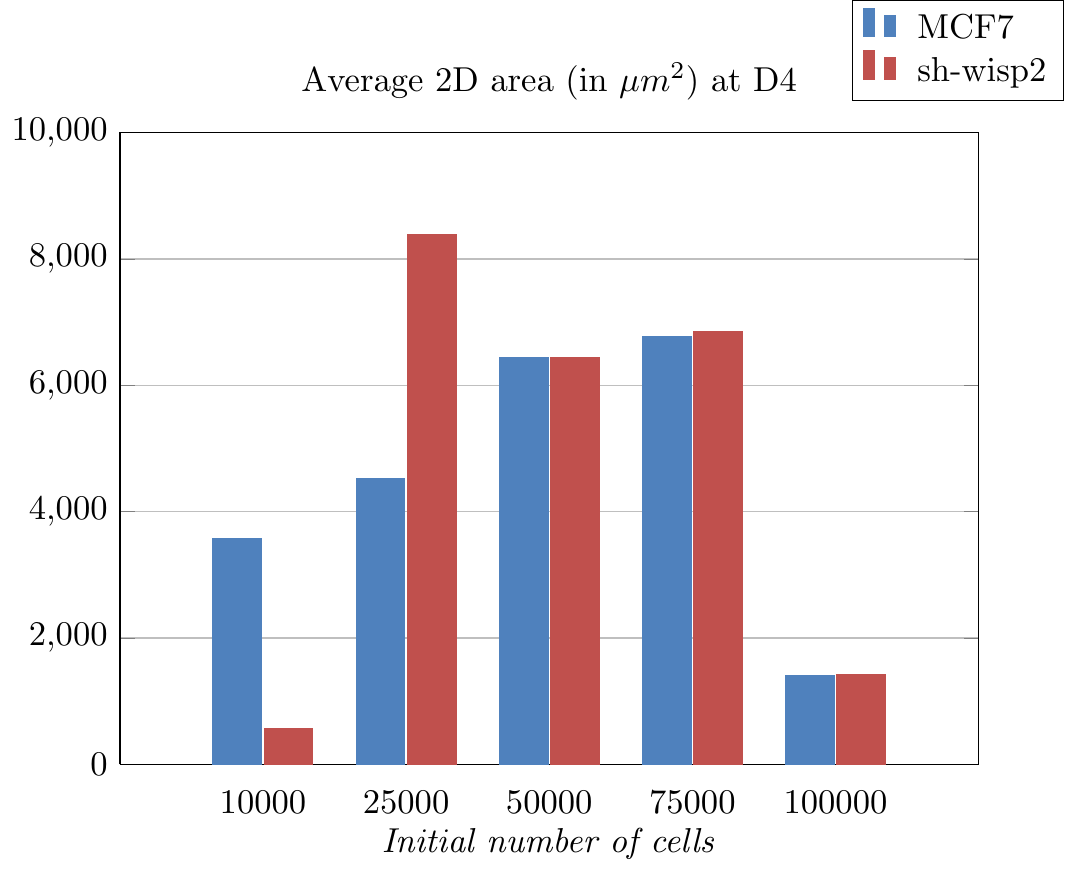}
		\caption{}
	\end{subfigure}
	\begin{subfigure}[]{0.3\textwidth}
		\includegraphics[scale=0.45]{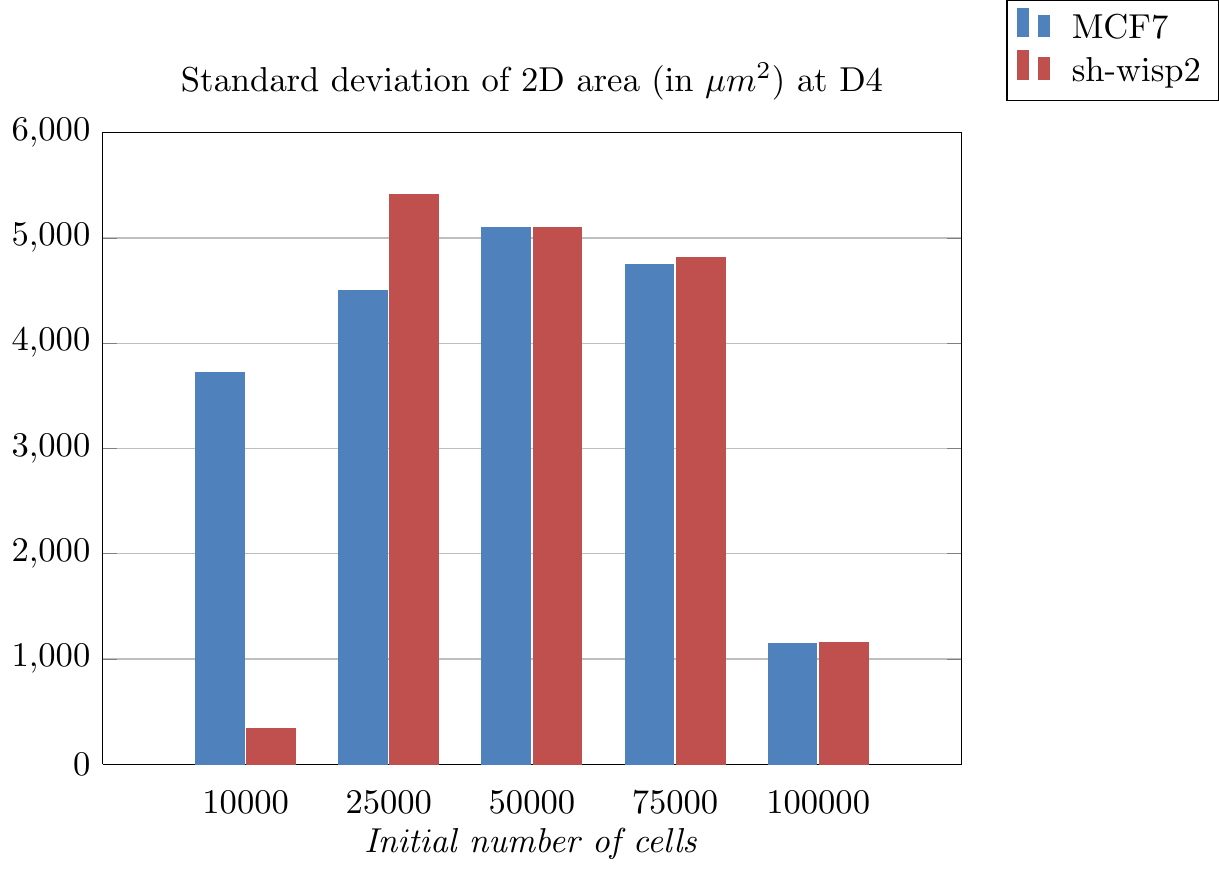}
		\caption{}
	\end{subfigure}
	\\[-2mm]
	\caption{Statistics of spheroids formed in the hydrogel by MCF7 and MCF7-sh-wisp2 cells after $4$ days of culture: number of spheres (A), mean and standard deviation of sizes of spheroids (B) and (C), both as a function of the number of seeded cells. Statistics obtained by averaging the results of $5$ randomly chosen images in a given hydrogel.}
	\label{Statistics}
\end{figure*}

A typical spheroid has a radius of around $100 \, \mu m$, and thus can be estimated to contain about $5\,000$ cells. As reported on Figure~\ref{Statistics}, at day D4 spheroids have appeared for both cell lines and for most number of seeded cells, which varies from $10\,000$ to $100\,000$. Interestingly, almost no spheroids are observed at low initial number of cells in the MCF7-sh-wisp2 case. We also note that the number of spheroids is about the same for $50\,000$ or $75\,000$ seeded cells. Finally, very few spheroids are observed at $100\,000$ initial cells. This last observation should be handled with care since it is essentially the effect of cells tending to escape and pack outside the hydrogel. Thus, $100\,000$ might be a too high initial number of cells for the experiment to run properly.

Building on these experiments, we propose that chemotaxis (along with diffusion and growth) is the driving force behind the emergence of these patterns. In our case, well-documented chemoattractants are the chemokines CXCL12 and CXCL8, mostly expressed by MCF7 cells and MCF7-sh-wisp2 cells respectively~\cite{Sabbah2011}. 

To support our claim, we provide a minimally-parametrized nonlinear variant of the Keller-Segel system. In agreement with the biological observation that cells have quickly spread uniformly, we consider a homogeneous initial condition, starting from which the system has an exponentially growing homogeneous time-dependent solution, driven by the growth of cells.

We investigate Turing instabilities for this model around the time-dependent solution. Assuming that the time-scale of growth is long compared to that of kinetics, we prove that such instabilities exist under a time-dependent necessary and sufficient condition, which we call $IC(t)$.
The instability condition provides us with a dynamical explanation of patterns:
\begin{itemize}
\item as long as $IC(t)$ is not satisfied, the solution evolves close to the homogeneous solution,
\item when $IC(t)$ starts being fulfilled, patterns arise on the quick time scale of kinetics.  
\end{itemize}

We discuss in detail and compare to experiments the dependence of $IC(t)$ on the initial number of cells and on the parameter for the invasiveness of cells. Most experimental results are shown to be coherent with the theoretical predictions, and they justify the nonlinearity chosen in the Keller-Segel system. Remaining mismatches are also analyzed and put in perspective with the minimality of our model. 

Finally, we provide 2D simulations for the equations in a disk, which both confirm our theoretical claims and show a qualitative match with experimental data. We prove the relevance of these 2D simulations when it comes to patterning, by highlighting that patterns in the 2D disk are an excellent approximation for their 3-dimensional counterpart in a cylinder, when the cylinder height is relatively small compared to its radius.  

\paragraph*{Materials and methods.} 

\begin{figure}[h!!]
	\centering
	\includegraphics[scale = 0.5]{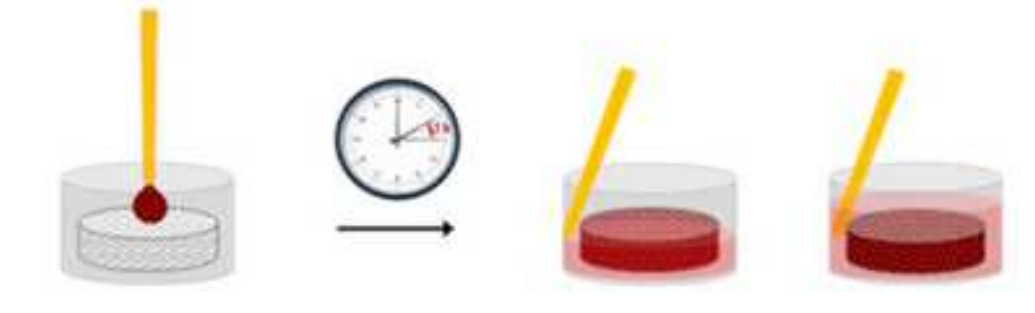}
	\caption{Experimental process of micro-tumor formation: cells are suspended on the top center of the hydrogel $96$-well-plates in a humidified atmosphere containing $5$\% CO2, at $37\degree$C. After $1$ hour of incubation, $150\,\mu l$ of medium is added in the space between the well and the hydrogel.}
	\label{fig:methode}
\end{figure}

A schematic view of the experimental process is given in Figure \ref{fig:methode}. It consists in the 3D culture of specific breast cancer cell lines using hydrogel $96$-low binding microwell arrays (Biomimesys) developed by Celenys. Cells counted and suspended in $50\, \mu l$ of medium were seeded at different densities, from $10\,000$ to $100\,000$ cells/well. Medium was changed every 2 days during the whole experiment, and when the spheres became too big, the hydrogels (containing spheres) were placed in $24$-well-plates to keep supplying nutrients to the cells.

Two breast cell lines were used for this study: MCF7 (human breast cell line, epithelial phenotype) and MCF7-sh-WISP2 (MCF7 cells invalidated for WISP2 by sh-RNA plasmid, mesenchymal phenotype, \cite{Ferrand2014, Fritah2008}). Cells were routinely maintained in Dulbecco's modified Eagle medium supplemented with $10$\% fetal bovine serum (FBS), L-Glutamine, and antibiotics.

Live-cell microscopy imaging of spheroids was obtained by using EVOS microscope (EVOS Cell Imaging System, Thermofisher). Images were collected with either x$4$ or x$10$ objectives depending on the sizes of spheroids after $4$ days of culture. Image analysis was carried out using the freely available ImageJ software. Spheroids were analyzed by using either defined (round or oval selection) or freehand selections.

\section{Mathematical model}
\label{sec:model}

The Keller-Segel system describes the directed motion of cells towards a chemo-attractant. We consider a well established variant which takes into account volume effects. The equation is posed on $\Omega$ which is either a cylinder for the hydrogel or a disk for simulations. We consider the system of reaction-diffusion partial differential equations that writes:
	\begin{align}
	\label{eq:Model}
	\begin{cases}
	\frac{\partial n}{\partial t} - D_1 \Delta n + \chi \nabla  \cdot \left(\varphi(n) \nabla c\right)  = r\, n,\\
	\frac{\partial c}{\partial t} - D_2 \Delta c = \alpha n - \beta c,
	\end{cases}
	\end{align}
where $n(t,x)$ is the density of the population of cancer cells and $c(t,x)$ is the concentration of the chemoattractant, for $x \in \Omega$ and~$t > 0$.

We impose no-flux boundary conditions at the boundary $\partial \Omega$
\begin{align*}
\begin{cases}
\left( D_1  \nabla n - \chi \varphi(n) \nabla c \right) \cdot \mathbf{n} = 0, \\
D_2  \nabla c  \cdot \mathbf{n} = 0,
\end{cases}
\end{align*}
where $\mathbf{n}$ is the outward unit normal at the boundary. With this choice, cells and chemoattractant neither leave nor enter the domain. We complete this system with an initial condition $n(0,x) = n^0(x)$ and $c(0,x) = c^0(x)$ and we define the average initial mass
\begin{equation*}
M := \frac{1}{|\Omega|}\int_\Omega n^0(x) \, \d x.
\end{equation*}

\subsection{Model description}
The second equation in~\eqref{sec:model} describes the dynamics of the concentration of the chemoattractant~$c(t,x)$: it diffuses at rate $D_2 > 0$, is produced by the cells themselves at rate $\alpha > 0$ and is naturally degraded at rate $\beta > 0$. 
Cells diffuse at a constant rate $D_1 > 0$ and grow according to a Lotka-Volterra law with intrinsic rate~$r$. Not only does this choice ease computations, but we argue it is an appropriate description of the phase before the emergence of patterns, since cells have much space to grow and nutrients are brought to them constantly.

Finally, the random diffusive movement is complemented with a biased divergence term in the direction of the gradient of chemoattractant~$c$. The constant $\chi > 0$ measures the strength of motion and $\varphi(n)$, called \textit{chemotactical sensitivity} function, is a nonlinear function describing the way cells aggregate when following the chemical signal.
We shall consider either
\begin{equation*}
\varphi(n) = n\left(1-\frac{n}{n_{\max}}\right)_+ \text{ or } \varphi(n) = n \, e^{-\frac{n}{n_{\max}}},
\end{equation*}
where $x_+$ stands for the positive part of a given real number $x$. These types of functions have been considered in various works such as~\cite{Hillen2000, Hillen2002}. We will show that a nonlinear choice of sensitivity is necessary for a correct agreement between theoretical predictions and experimental results. 

\begin{remark}
Because of the linear growth for cells, the density $n_{\max}$ will eventually be exceeded, in which case a nonlinearity $\varphi(n) = n(1-\frac{n}{n_{\max}})$ would not make sense since cells would go in the direction given by $-\nabla c$. This explains our choice with the positive part which completely shuts down any chemotactic movement above the maximal density $n_{\max}$.
\end{remark}

\noindent Since biological experiments have been performed in 3D cylinders of small height, the first simplification we propose is to consider the problem \eqref{eq:Model} in a disk, thus with $\Omega \subset \mathbb{R}^2$. As we shall prove rigorously in Appendix~\ref{AppA}, from the point of view of pattern formation we are interested in, we can neglect the third dimension. In fact, when the height of the cylinder is small, the perturbations along the $z-$axis will not be observed.

As observations show, cells quickly spread uniformly in the 3D structure and we start after this spreading phase: we assume
\begin{align*}
n^0(x) \equiv M, \quad c^0(x) \equiv  \frac{\alpha}{\beta} M,
\end{align*}
where $M$ is defined above and represents (up to the scaling factor $\frac{1}{|\Omega|}$) the initial number of cells seeded in the 3D structure.

\subsection{Dimensionless model}
Upon changes of time and space variables $\tilde{t} = \beta \frac{D_1}{D_2} t$, $\tilde{x} = \sqrt{\frac{\beta}{D_2}} x$ and appropriate scalings for $n$ and $c$, namely 
\begin{equation}
\begin{split}
n(t,x) =& \tilde{n} \left(\beta \frac{D_1}{D_2} t, \sqrt{\frac{\beta}{D_2}} x \right), \\
c(t,x) =& \frac{\alpha}{\beta} \tilde{c} \left( \beta \frac{D_1}{D_2} t, \sqrt{\frac{\beta}{D_2}} x \right),
\end{split}
\end{equation}
and writing again $n$ for $\tilde{n}$, $c$ for $\tilde{c}$ we find a minimally parametrized version: 
\begin{equation}
\label{P4C1ModelDimensionless}
\begin{cases}
\frac{\partial n}{\partial t} -  \Delta n + A \nabla  \cdot \left(\varphi(n) \nabla c \right) = r_0 \, n,\\
\e \frac{\partial c}{\partial t} -  \Delta c = n - c.
\end{cases}
\end{equation}
Only three parameters $A$, $\e$ and $r_0$ now remain, given by 
\begin{equation*}
A = \frac{\alpha \chi}{\beta D_1}, \; \e = \frac{D_1}{D_2}, \; r_0=\frac{r}{\beta \e}.
\end{equation*}
We expect $A$ to be larger for MCF7-sh-wisp2 cells than for MCF7 cells, the former being more prone to chemotactic movement than the latter~\cite{Fritah2008, Sabbah2011}.

\paragraph{Small parameters.}
Because the chemoattractant diffuses much faster than cells, $\e$ is thus typically small, while $A$ depends on the ratio $\frac{\chi}{D_1}$, which measures the relative importance of diffusion and attraction. 

We also assume that the time scale of growth (driven by $r$) is much bigger than the time scale of kinetics (driven by $\alpha$, $\beta$). Thus, $r_0$ is much smaller than $\frac{1}{\e}$. The MCF7 cells indeed have a population doubling time of around $1$ day~\cite{Sutherland1983}, which yields $r \approx 4. 10^{-6} \, s^{-1}$, while some data can be found in the literature on the degradation rate of the chemokine CXCL8, of the order of $\beta \approx 1. 10^{-4} \, s^{-1}$~\cite{Shi1995}. 

Summing up, we assume both 
\begin{equation}
\e  \ll 1, \qquad r_0 \, \e = \frac{r}{\beta}  \ll 1.
\end{equation}

For the homogeneous initial condition $n^0=M$, $c^0 = M$, this system has a homogeneous (in space) solution, given by 
\begin{equation*}
\bar{n}(t) := M e^{r_0 t},  \; \; \bar{c}(t) := \frac{M}{1+\e r_0}\left(\e r_0 e^{-\frac{t}{\e}}+e^{r_0t}\right),
\end{equation*}
the (linear) stability of which we now investigate in detail. 

\section{Linear stability analysis}
\label{sec:analysis}

Around the homogeneous solution $(\bar{n}(t), \bar{c}(t))$, the linearized system reads
\begin{equation}
\label{P4C1ModelLinearised}
\begin{cases}
\frac{\partial n}{\partial t} - \Delta n +  A \varphi(\bar{n}(t)) \Delta c = r_0\, n, \\
\e \frac{\partial c}{\partial t} - \Delta c = n - c, 
\end{cases}
\end{equation}
%%%%
We denote $(\psi_k)_{k \geq 1}$ the orthonormal basis of $L^2(\Omega)$ made of the eigenfunctions of the Neumann Laplace operator associated with eigenvalues $(\lambda_k)_{k \geq 1}$, namely 
\begin{equation*}
\begin{cases}
- \Delta \psi_k = \lambda_k \psi_k, \\
\frac{\partial \psi_k}{\partial \nu} = 0. 
\end{cases}
\end{equation*}
%%%%
Projecting the linearized equation~\eqref{P4C1ModelLinearised} on the orthonormal basis $(\psi_k)_{k \geq 1}$ through \begin{equation*}
n(t,\cdot) = \sum a_k(t) \psi_k, \; c(t,\cdot) = \sum b_k(t) \psi_k,
\end{equation*}
we find
\begin{equation*}
\begin{cases} 
a_k'(t) = -\lambda_k a_k(t) + A \varphi(\bar{n}(t)) \lambda_k b_k(t) + r_0 a_k(t), \\
\e b_k'(t) = -\lambda_k b_k(t) + a_k(t) - b_k(t).
\end{cases}
\end{equation*}
%%%%

The previous equation writes in matrix form as $X'_k(t) = A_k(t) X_k(t)$ with
\begin{equation*}
X_k(t) = \begin{pmatrix}
a_k(t) \\
b_k(t) \\
 \end{pmatrix},  \;A_k(t)=
  \begin{pmatrix}
-\lambda_k + r_0 & A \varphi(\bar{n}(t)) \lambda_k \\
 \frac{1}{\e}  & -\frac{1}{\e}(\lambda_k +1) \\
 \end{pmatrix}.
\end{equation*}

We now assume that the perturbation is initiated at time $t_0>0$ and we focus on an interval of the form $(t_0, t_0 + \Delta t)$ whose size $\Delta t$ is of the order of the kinetics time-scale~$\e$. As such, it is small when compared to the growth time-scale $\frac{1}{r_0}$. Note that in the original time variable, this amounts to considering a time interval of order $\frac{1}{\beta}$ (small when compared to $\frac{1}{r}$). In particular, neglecting terms of order $\mathcal{O}((r_0 \e))$, we can approximate all functions of time by their value at $t_0$.

Looking for exponentially increasing solutions in time, we insert a solution of the form 
\begin{equation*}
(a_k(t), b_k(t)) = e^{\mu (t-t_0)} (a_k^0, b_k^0),
\end{equation*}
with $\mu$ of real part $\Re(\mu) >0$, which imposes that $\mu$ is an eigenvalue of $A_k(t_0)$). Let us denote $\mu_k^+(t)$ to be either the largest real eigenvalue of $A_k(t)$ or the real part of its complex conjugates eigenvalues.

We now look for sufficient and necessary conditions ensuring that $\mu_k^+(t)>0$.
 
For a given time $t >0$, we compute 
\begin{equation*}
\text{Tr}(A_k(t)) = - \lambda_k + r_0 - \frac{1}{\e}(\lambda_k +1) \leq r_0 - \frac{1}{\e} <0.
\end{equation*}
It is thus easy to check that $A_k(t)$ has an eigenvalue with positive real part if and only if $\det(A_k(t)) < 0$, which is equivalent to 
\begin{equation*}
-\lambda_k^2 + \left(A \varphi(\bar{n}(t)) + r_0 -1\right) \lambda_k + r_0 > 0.
\end{equation*} 

Since we are interested in perturbations other than those in the direction of the first homogeneous eigenfunction $\psi_1$ (and since the above polynomial is positive at $0$), a necessary and sufficient condition to have $\mu_k^+(t)>0$ is for $\lambda_k>0$ to satisfy 
\begin{equation*}
\lambda_k  < \bar \lambda(t)
\end{equation*}

where 
\begin{equation*}
 \textstyle \bar \lambda(t) := \frac{r_0 + A\varphi(\bar{n}(t)) -1 + \sqrt{(r_0 + A\varphi(\bar{n}(t)) -1)^2 + 4r_0}}{2}. 
\end{equation*}

In other words, looking for perturbations in a direction other than that of the eigenfunction $\psi_1$ which is homogeneous, a perturbation at time $t_0$ will yield Turing instability if and only if
\begin{equation}
\label{TuringInstability}
\lambda_2  < \bar \lambda(t_0), 
\end{equation}
neglecting small terms in $\mathcal{O}((r_0 \e))$. As in the Introduction, we call this condition $IC(t_0)$. 

\begin{remark}
If one replaces $\bar \lambda(t_0)$ by $\bar \lambda(t_0 + \frac{\Delta t}{2})$ in the previous condition, the instability condition can be proved to be accurate not only at order $0$ but even at order $1$ in $r_0 \, \e$, \textit{i.e.}, neglecting $\mathcal{O}((r_0 \e)^2)$ corrections, see~\cite{Madzvamuse2010}.
\end{remark} 
%Recalling the smallness assumption on $r_0$, a simpler approximate condition is 
%\[\lambda_2 +1 <  A\varphi\bigg(\bar{n}\Big(\frac{t_0+\Delta t}{2}\Big)\bigg).\]
The perturbations happen along the modes $\psi_k$ which for simple geometries as in our case can be explicitly computed, see Appendix~\ref{AppA}.
\paragraph{On the condition for Turing instability.}
In condition~\eqref{TuringInstability}, the right-hand side $\bar \lambda(t)$ has the same monotonicity as $\varphi$ as a function of $\bar{n}(t_0) = M e^{r_0 t_0}$. Recall that the function $\varphi$ increases and then decreases in both cases $\varphi(n)= n(1-\frac{n}{K})_+$ and $\varphi(n) = ne^{-\frac{n}{K}}$. 
 
 \textit{Dependence on time.}\; 
 Assume that $M$ is fixed such that initially, $\varphi(\bar{n}(t))$ increases with time (\textit{i.e.}, $\varphi'(M)>0$). Consequently, the condition $IC(t)$ in~\eqref{TuringInstability} might not initially be satisfied but it is more likely to be as time increases. The typical dynamics expected from this condition is thus an increase of $n(t,\cdot)$, $c(t, \cdot)$ very close to $\bar{n}(t)$, $\bar{c}(t)$, up until the Turing instability condition becomes satisfied. Patterns then form very quickly (on the time scale of $\e$). 
 
\textit{Dependence on the initial mass.}\; 
For a small initial mass $M$, the right-hand side in condition $IC(t)$ in~\eqref{TuringInstability} is small, meaning that Turing instabilities are expected only at a large time $t$. This provides an explanation for the fact that patterns are not initially observed. In the limit when $M$ is very small, one should wait for a very long time before $IC(t)$ starts being satisfied.

As $M$ increases, Turing instability is more likely to occur before D4, until we reach the cell density reaches the region in which $\varphi$ decreases. When $M$ is too large, $IC(t)$ is not satisfied initially and will never be later on: no patterns should be obtained. This in accordance with experimental observations, and explains why we choose a nonlinear sensitivity function instead of the more classical linear one. A linear choice would indeed predict Turing instabilities for high values of $M$.

\textit{Dependence on the parameter $A$.}\; 
We note that the dependence of $IC(t)$ in~\eqref{TuringInstability} as a function of $A$ is such that if $IC(t)$ is satisfied at some time $t$ for a given $A$, then it should be for a larger $A$ as well. In other words, if MCF7 cells have created patterns at day D4, then so should have the MCF7-sh-wisp2 (all other parameters being equal). This is not in agreement with the experimental findings with $M = 10 000$, value for which MCF7 cells have created spheroids, but not MCF7-sh-wisp2. We postulate this comes from usual shortcomings of deterministic PDE models at low densities.
\section{Numerical simulations}
\label{sec:simulations}

\begin{figure*}[h!!]
	\centering
	\includegraphics[width=\textwidth]{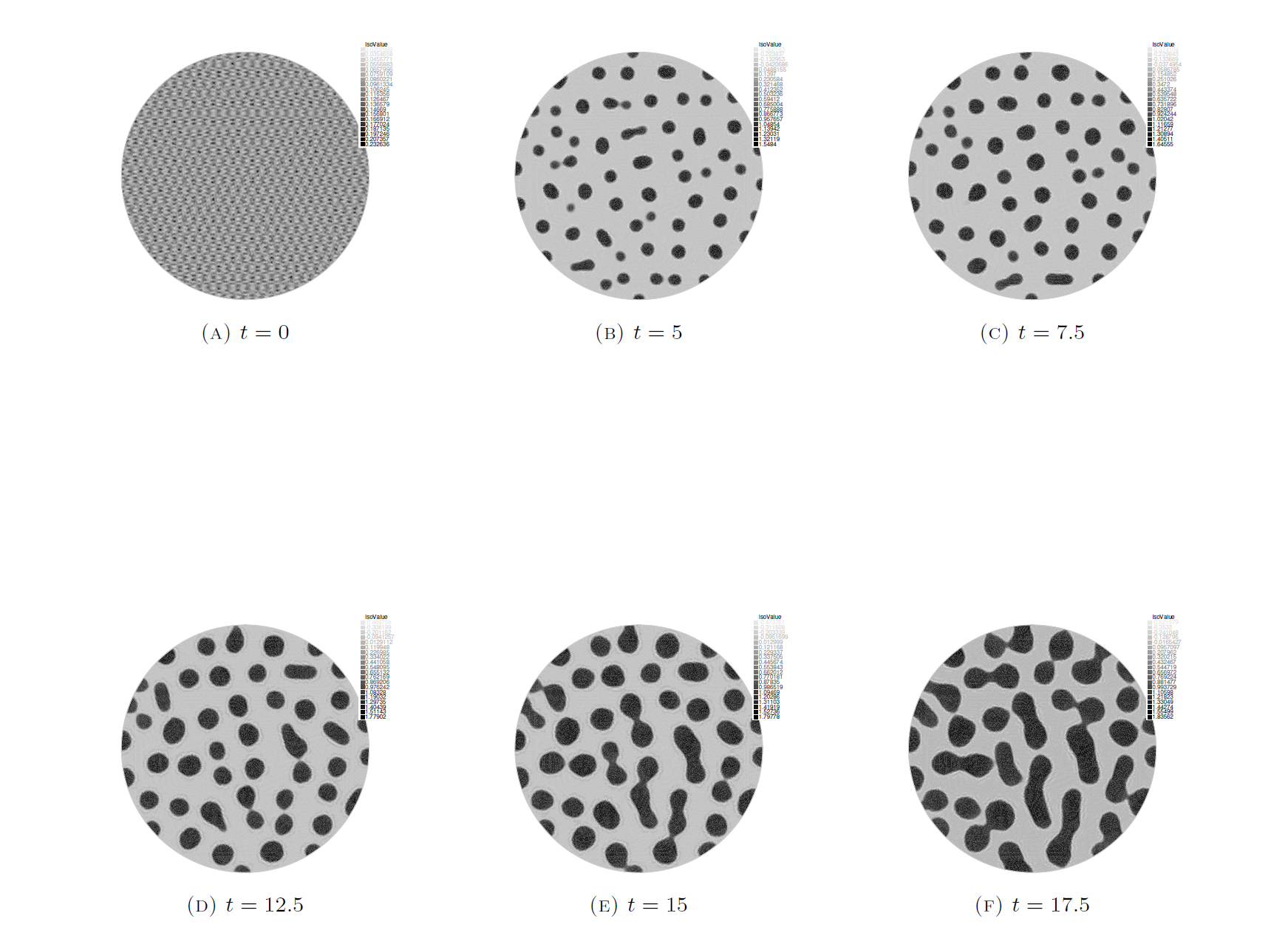}
	\caption{\textsc{Patterning evolution}. Evolution of the density of cells from a small perturbation of the initial uniform state $n_0 = 0.1$ ($t = 0$). Here, $\varphi(n) = n\left(1-n\right)_+$, $A = 70$, $\varepsilon = 0.01$ and $r_0 = 0.1$.}
\label{fig:patterns_evolution}
\end{figure*}

\begin{comment}
\begin{figure*}[h!!]
	\centering
	\begin{subfigure}[b]{0.3\textwidth}
		\includegraphics[width=\textwidth]{case1_logistic_growth_t0.pdf}
		\caption{$t = 0$}
		\label{fig:logistic_SG_UP_50}
	\end{subfigure}
	\begin{subfigure}[b]{0.3\textwidth}
		\includegraphics[width=\textwidth]{case1_logistic_growth_t5.pdf}
		\caption{$t = 5$}
	\end{subfigure}
	\begin{subfigure}[b]{0.3\textwidth}
		\includegraphics[width=\textwidth]{case1_logistic_growth_t7point5.pdf}
		\caption{$t = 7.5$}
	\end{subfigure}
	\\[-2mm]
	\begin{subfigure}[b]{0.3\textwidth}
		\includegraphics[width=\textwidth]{case1_logistic_growth_t12point5.pdf}
		\caption{$t = 12.5$}
	\end{subfigure}
	\begin{subfigure}[b]{0.3\textwidth}
		\includegraphics[width=\textwidth]{case1_logistic_growth_t15.pdf}
		\caption{$t = 15$}
	\end{subfigure}
	\begin{subfigure}[b]{0.3\textwidth}
		\includegraphics[width=\textwidth]{case1_logistic_growth_t17point5.pdf}
		\caption{$t = 17.5$}
	\end{subfigure}
	\caption{\textsc{Patterning evolution}. Evolution of the density of cells from a small perturbation of the initial uniform state $n_0 = 0.1$ ($t = 0$). Here, $\varphi(n) = n\left(1-n\right)_+$, $A = 70$, $\varepsilon = 0.01$ and $r_0 = 0.1$.}
	\label{fig:patterns_evolution}
\end{figure*}
\end{comment}

In this section, we complete the theoretical analysis with numerical simulations of system \eqref{P4C1ModelDimensionless} showing how the patterning behavior of solution changes depending only on few parameters, as commented in the previous sections. All simulations have been performed with the software Freefem++, which is based on finite element methods~\cite{Hecht2012}. They have been carried out on a circle domain of radius $R = 20$, discretized with a $500$ points mesh.

 \textit{Typical behavior.}\; 
When the parameters are in the instability region \eqref{TuringInstability}, the typical behavior of the solutions is the one showed in \Cref{fig:patterns_evolution}, where we set $A = 70$, $\varepsilon = 0.01$, $r_0=0.1$ and used the logistic chemotactical sensitivity function, \textit{i.e.}, $\varphi(n) = n\left(1-n\right)_+$. Starting from a small perturbation of the constant initial distribution $n_0 = 0.1$, in few time steps the density grows uniformly in the domain until $IC(t)$ in~\eqref{TuringInstability} is satisfied, in this case around $t \approx 2.5$. Pattern formation then occurs very quickly, compared to the growth dynamics. When patterns are formed, the solution evolves very slowly and the most visible phenomenon is the merging of the spheroidal aggregates. Eventually, they become less regular, round structures, as at $t = 17.5$.

\begin{figure}[h!!]
	\centering
 	\includegraphics[width=0.3\textwidth]{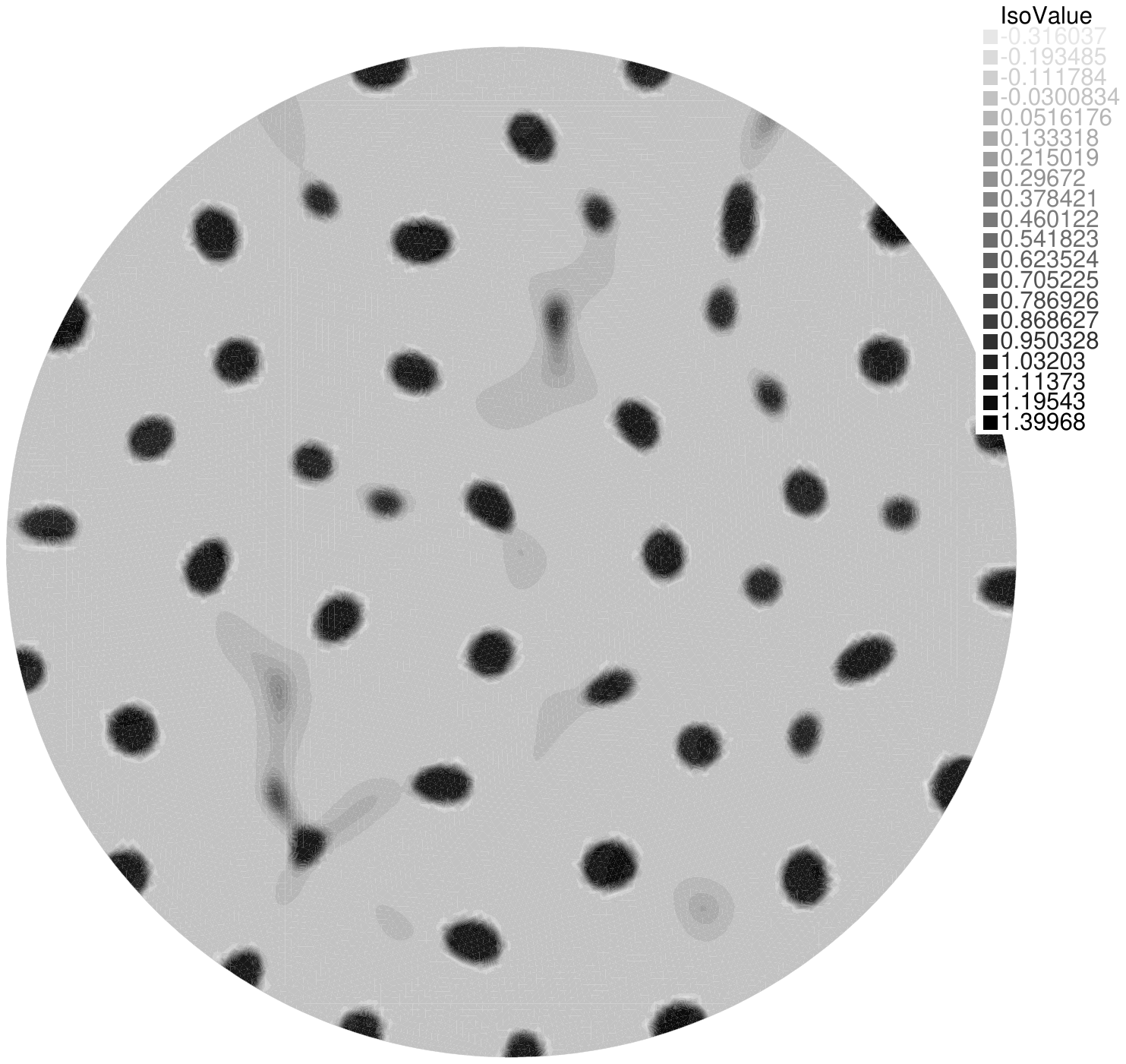}
	\caption{Patterns appear around time $t = 7.5$ when $n_0 = 0.05$, $A = 70$, $\varepsilon = 0.01$ and $r_0 = 0.1$.}
	\label{fig:initial_condition}
\end{figure}

\textit{Dependence on the initial mass.}\; 
If the initial mass is too small, the initial phase of growth until takes much longer and the patterns arise later. For example, in~\Cref{fig:initial_condition} where the solution is displayed with the initial distribution value~$n_0 =  0.05$ (leaving all the other parameters unchanged), the first patterns appear only around~$t \approx 7.5$. 

\begin{figure}[h!!]
	\centering
	\begin{subfigure}[b]{0.3\textwidth}
		\includegraphics[width=\textwidth]{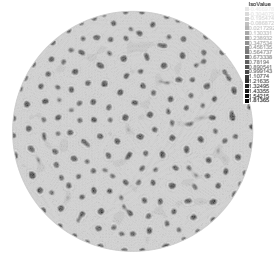}
		\caption{$t = 10$}
	\end{subfigure}
	\begin{subfigure}[b]{0.3\textwidth}
		\includegraphics[width=\textwidth]{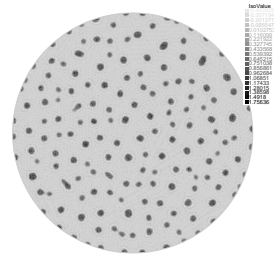}
		\caption{$t = 17.5$}
	\end{subfigure}
	\\[-2mm]
	\caption{Evolution of the density of cells from a small perturbation of the initial uniform state $n_0 = 0.1$, with $\varphi(n) = n\left(1-n\right)_+$, $A = 200$, $\varepsilon = 0.1$ and $r_0 = 0.1$.}
	\label{fig:patterns_evolution2}
\end{figure}

\textit{Dependence on the parameter $A$.}\; 
In~\Cref{fig:patterns_evolution2}, we show how the kind of patterns observed strongly depends on the value of the quantity $\chi / D_1$. In these simulations, we chose $A = 200$, $\varepsilon = 0.1$ and the other parameters as in \Cref{fig:patterns_evolution}. In this case the diffusivity of cells is not as strong and the chemotactical attraction dominates the dynamics, leading to smaller but more numerous spheroids. Moreover, since the merging of aggregates is now less frequent, slightly more non-spheroidal, elongated patterns arise. 

\begin{figure}[h!!]
	\centering
	\begin{subfigure}[b]{0.3\textwidth}
		\includegraphics[width=\textwidth]{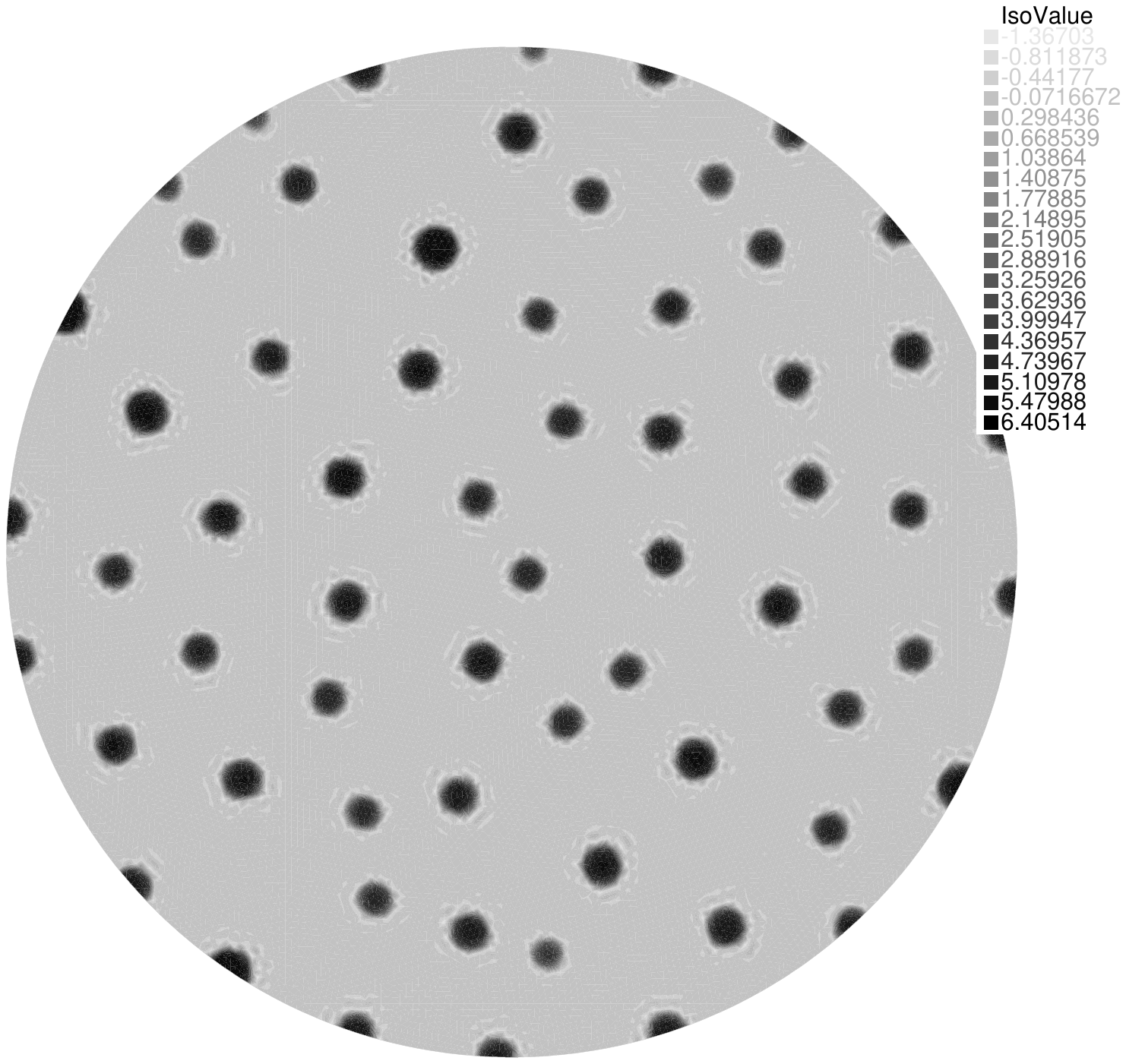}
		\caption{$t = 12.5$}
	\end{subfigure}
	\begin{subfigure}[b]{0.3\textwidth}
		\includegraphics[width=\textwidth]{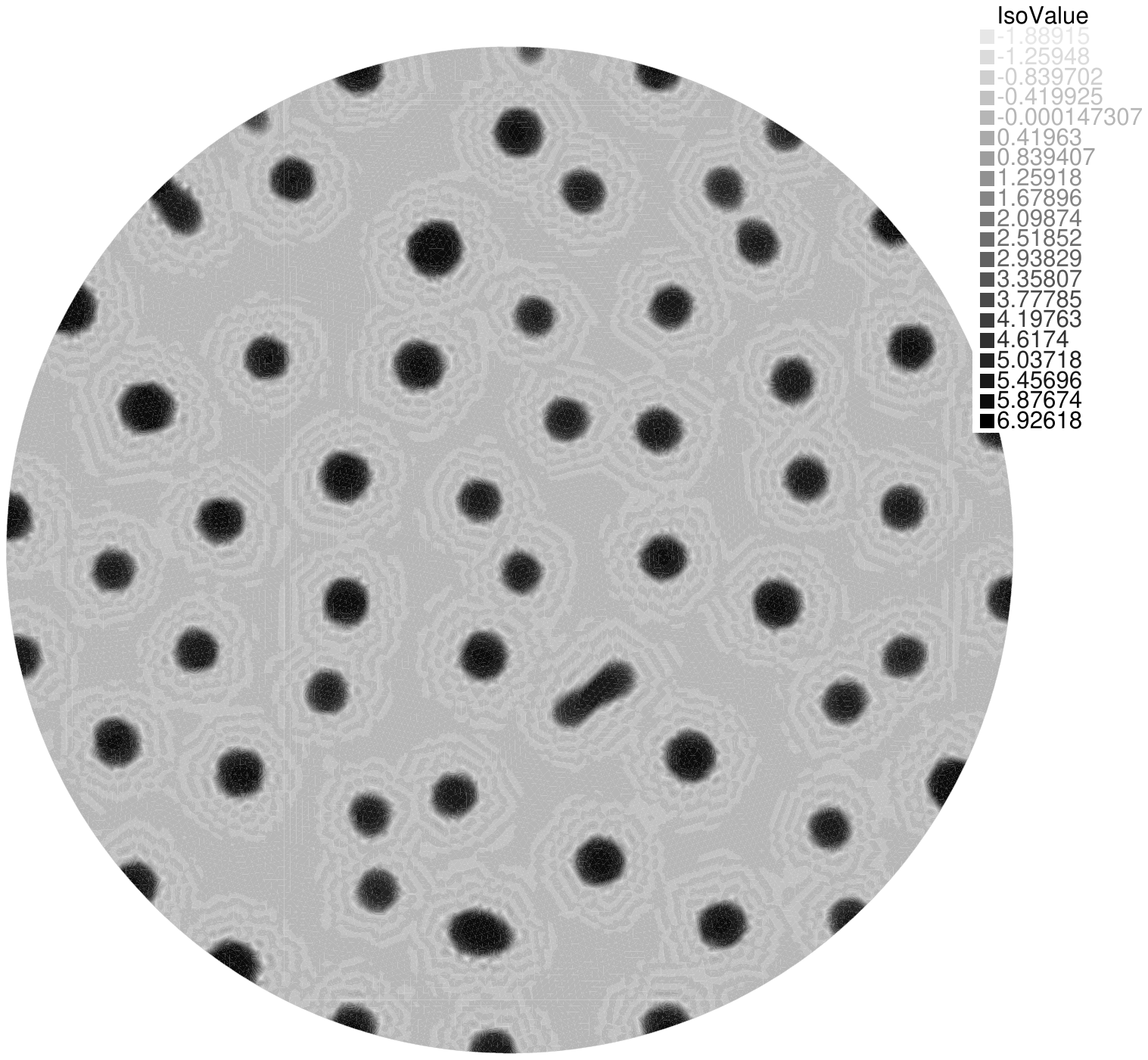}
		\caption{$t = 17.5$}
	\end{subfigure}
	\\[-2mm]
	\caption{\textsc{Alternative model}. Evolution of the density of cells from the initial state $n_0 = 0.1$ in the case where $\varphi(n) = n e^{-n}$. All parameters have been chosen as in \Cref{fig:patterns_evolution}.}
	\label{fig:patterns_exp}
\end{figure}

\textit{Choice of the function $\varphi$.}\; 
Finally, \Cref{fig:patterns_exp} shows solutions of the system \eqref{P4C1ModelDimensionless} with $\varphi(n) = n e^{-n}$. For this chemotactical sensitivity function, we find again spheroidal aggregates, but with less variable structures: in this case, variability lies in the maximums of the solution (higher than the ones in the logistic case), but only very round aggregates can be observed.

While the exponential function is usually advocated for because it does not impose an a priori maximal density, we here highlight a drawback of this choice: since the packing can continue even at high density, this fixes the size of patterns as new cells do not go at the periphery but instead concentrate at higher and higher density at the middle. Thus, this choice of function offers less variability for the size of the spheroids.

Note that the numerical schemes do not preserve desirable properties, such as positivity of the solution. In fact, negative values in the densities can be seen mostly at the interfaces between zones of high densities (the spheroids), and zones of low densities (the rest). We were able to take fine enough grids to limit these numerical artifacts but we point out that some positivity-preserving numerical schemes have been designed to solve systems of the type of~\eqref{eq:Model}, see for example \cite{Almeida2018} and \cite{LiuWangZhou2018}.

\section{Discussion and conclusions}
\label{sec:conclusions}
In order to understand the experimental observation of spheroidal aggregates in cultures of breast cancer cells, we have proposed a mathematical model which includes a chemotactic effect and its modulation at high cell densities. We have analyzed theoretically the conditions which lead to instability and pattern formation. The numerical  olutions confirm the theoretical analysis and  qualitatively reproduce the observed patterns, in spite of its relatively low
parametrization. Contrarily to other modeling approaches, the salient feature is not cell-scaffold adhesion, which we reduce to a constant diffusion term, but instead chemotaxis. We thus hypothesize that it is a key phenomenon responsible for these aggregates.

 The numerical simulations of the model do not only show qualitative accordance with the experimental results. Indeed, the system~\ref{eq:Model} has spatially inhomogeneous solutions with spheroidal patterns, and it also describes how different kinds of patterns can arise: few, elongated structures for a small diffusion value or numerous, mainly round, small aggregates for a stronger chemotactic sensitivity.

The simplicity of the model induces two main limitations:
\begin{itemize}
	\item it does not seem to offer a satisfying flexibility for the size of patterns, essentially fixed by the geometry of $\Omega$. 
	\item it is not suitable to explain the post-aggregation phase, and in particular the increase in the number of spheres as observed experimentally 10 days after initial pattern formation.
\end{itemize}
In fact, the distributions (in size) of spheroids for the biological images and the numerical simulations do not match well: while the standard deviation is of the order of the average size for experiments as evidenced by Figure~\ref{Statistics} in the Introduction, we find that standard deviation is about one third of the mean size in simulations. 

In the Keller-Segel model, the variability is essentially captured only by the Laplacian eigenfunctions which themselves are completely characterized by the domain geometry.  A natural direction of research for a better matching is to model cell-scaffold adhesion more finely than with a diffusion term, incorporating anisotropies (such as in~\cite{Painter2009}), or even randomness, in the extracellular matrix density.

As for the second point, statistical estimates obtained from images taken later during the experiment (not shown here) indeed evidence a growth in both the size and number of spheroids. This is not reproduced by numerical simulations. In fact, the patterns that formed then typically continue to merge, probably until the cells are all packed in very few aggregates. This phenomenon for this type of model is explained in detail in~\cite{Potapov2005}. 

We insist that a minimally-parametrized model such as ours is more amenable to mathematical analysis and also paves the way for works aiming at a more quantitative prediction of the typical size and number of spheroids. To go further in this direction, one should look for the actual modes along which instabilities will be observed in a time-dependent setting, in the spirit of~\cite{Madzvamuse2010}.
\\[3mm]
{\bf Acknowledment.}  
The authors acknowledge partial funding from the ANR blanche project Kibord ANR-13-BS01-0004 funded by the French Ministry of Research. 
\\
B.P. has received funding from the European Research Council (ERC) under the European Union's Horizon 2020 research and innovation programme (grant agreement No 740623).

\appendix
\section{Explicit computation of the modes}
\label{AppA}
Since $\Omega$ has a particular shape, eigenvalues and eigenfunctions can actually be explicitly computed. 
We first consider the case of the 2D simulations, namely when $\Omega$ is a disk of radius $a$. It is then standard that all eigenfunctions can be obtained after separation of variables in polar coordinates $\psi(x,y) = f(r) g(\theta)$, the equation $-\Delta \psi = \lambda \psi$ with Neumann boundary conditions is equivalent to 
\begin{equation*}
g(\theta) = A \cos(m\theta) + B \sin(m\theta)
\end{equation*}
for some $m \in \mathbb{Z}$ and $\rho \mapsto f(\frac{\rho}{\sqrt{\lambda}})$ must solve the Bessel equation
\begin{equation*}\rho^2 y''(\rho) + \rho y'(\rho) + (\rho^2 - m^2) y(\rho) = 0
\end{equation*}
with $y'(0) = y'(\sqrt{\lambda}a) = 0$. 
This yields, up to a constant, to the result $f(r) = J_m (\sqrt{\lambda} r)$ where $J_m$ is the first kind Bessel function of order $m$. The boundary conditions impose $m \neq \pm 1$ (because $J_m'(0) = 0$ for all $m$ except $1$ and $-1$), while, denoting $\gamma_{m,p}$ the $p$th zero of the derivative of $J_m$, we find $\lambda = \big(\frac{\gamma_{m,p}}{a}\big)^2$.

Summing up, we obtain 
\begin{align*}\lambda_{m,p} & =\left(\frac{\gamma_{m,p}}{a}\right)^2, \\ \psi_{m,p}(r, \theta) & = J_m\left(\frac{\gamma_{m,p}}{a}r\right) (A  \cos(m\theta) + B \sin(m\theta)),\end{align*}
a family indexed by $m \in \mathbb{Z}  \setminus\{ \pm 1\}$, $p \in \mathbb{N}^\star$. Apart from $m=0$ for which the eigenfunction is unique after normalisation, the eigenspace associated to $\lambda_{m,p}$ is of dimension $2$.

Similar computations for the case of a cylinder of height $h$ and radius $a$ lead to the result 
\begin{align*} 
\lambda_{m,p,l} & =\left(\frac{\gamma_{m,p}}{a}\right)^2 + \left(\frac{l \pi}{h} \right)^2, \\
\psi_{m,p,l}(r, \theta,z ) & =\psi_{m,p}(r, \theta)  \cos\left(\frac{l \pi z}{h}\right),
\end{align*}
a family indexed by $m \in \mathbb{Z}  \setminus \{\pm 1\}$, $p \in \mathbb{N}^\star$, $l \in \mathbb{N}$. The multiplicity of eigenfunctions is the same as in the previous case ($1$ if $m=0$ and $2$ if not). 

If $h$ is small, the contribution of the $(\frac{l \pi}{h})^2$ term is too big and we will thus typically not see the eigenvalues such that $l>0$ in the $z$ variable, an approximation precise up to $\mathcal{O}((\frac{a}{h})^2)$ errors.  As a consequence, only the corresponding modes for $l=0$ will be observed, and we note that these are exactly the 2D modes.

An interesting and relevant consequence is that from the point of view of Turing instabilities, it is a good approximation to neglect the $z$ variable and focus on $\Omega \subset \mathbb{R}^2$ as a disk for simulations.

{%
\scriptsize

	\bibliography{bibliography}
	
	\bibliographystyle{acm}}

\end{document}